\documentclass[lettersize,journal]{IEEEtran}
\usepackage{amsmath,amsfonts}
\usepackage{algorithmic}
\usepackage{algorithm}
\usepackage{array}
\usepackage[caption=false,font=normalsize,labelfont=sf,textfont=sf]{subfig}
\usepackage{textcomp}
\usepackage{stfloats}
\usepackage{url}
\usepackage{verbatim}
\usepackage{graphicx}
\usepackage{cite}
\usepackage{lipsum}  

\usepackage{xcolor}

\newcommand{\quant}[1]{{\textcolor{teal}{#1}}}
\newcommand{\quantb}[1]{{\textcolor{violet}{\textbf{#1}}}}

\usepackage[acronyms,toc,nonumberlist,nogroupskip]{glossaries}
\glsdisablehyper
\glsnoexpandfields
\newacronym{ads}{ADS}{Automated Driving System}
\newacronym{hmi}{HMI}{Human Machine Interface}
\newacronym{odd}{ODD}{Operational Design Domain}
\newacronym{fzd}{FZD}{Institute of automotive engineering Darmstadt}
\newacronym{iad}{IAD}{Institute of Ergonomics and Human Factors Darmstadt}
\newacronym{acc}{ACC}{Adaptive Cruise Control}
\newacronym{sd}{SD}{Standart Deviation}
\newacronym{woz}{WOz}{Wizard of Oz}
\newacronym{ueq}{UEQ}{User Experience Questionnaire}
\newacronym{oedr}{OEDR}{Object and Event Detection and Response}

\begin{document}

\title{Acceptance and Trust: Drivers' First Contact with Released Automated Vehicles in Naturalistic Traffic}

\author{Sarah Schwindt-Drews\thefootnote{*}, Kai Storms\thefootnote{*}, Steven Peters, Bettina Abendroth
}



\maketitle

\def\thefootnote{*}\footnotetext{The authors contributed equally to this work.}

\begin{abstract}
This study investigates the impact of initial contact of drivers with an SAE Level 3 \gls{ads} under real traffic conditions, focusing on the Mercedes-Benz Drive Pilot in the EQS. 
It examines Acceptance, Trust, Usability, and User Experience. Although previous studies in simulated environments provided insights into human-automation interaction, real-world experiences can differ significantly. 
The research was conducted on a segment of German interstate with 30 participants lacking familiarity with Level 3 \gls{ads}. 
Pre- and post-driving questionnaires were used to assess changes in acceptance and confidence. Supplementary metrics included post-driving ratings for usability and user experience. 
Findings reveal a significant increase in acceptance and trust following the first contact, confirming results from prior simulator studies. 
Factors such as Performance Expectancy, Effort Expectancy, Facilitating Condition, Self-Efficacy, and Behavioral Intention to use the vehicle were rated higher after initial contact with the \gls{ads}. 
However, inadequate communication from the \gls{ads} to the human driver was detected, highlighting the need for improved communication to prevent misuse or confusion about the operating mode. 
Contrary to prior research, we found no significant impact of general attitudes towards technological innovation on acceptance and trust.
However, it's worth noting that most participants already had a high affinity for technology.
Although overall reception was positive and showed an upward trend post first contact, the \gls{ads} was also perceived as demanding as manual driving. 
Future research should focus on a more diverse participant sample and include longer or multiple real-traffic trips to understand behavioral adaptations over time.

\end{abstract}


\section{Introduction}
\IEEEPARstart{W}{ith} the introduction of automated vehicles, drivers will be able to travel more comfortable in the future \cite{Weigl.2022,Lehtonen.2022}. 
Drivers in vehicles with SAE Level~3 are currently authorized to temporarily shift their focus from driving to other non-driving related tasks (NDRTs) \cite{SAEInternational.04.2021}.
Nevertheless, they must always be capable of resuming the driving task when requested to do so by the \gls{ads}. 
Automated vehicles have the potential of increasing road safety by minimizing human error \cite{Nastjuk.2020}. 
Still, drivers must use these systems properly to achieve these benefits. 
A more comfortable and safer journey is only achievable if drivers accept the automated systems accordingly in order to actually use the functions \cite{Beggiato.2013,Korber.2018}. 
Adequate trust also is a critical variable for the appropriate use of automated vehicles \cite{Beggiato.2013,Korber.2018}. 
A lack of trust results in the non-utilization of the functions. Conversely, overtrust can be hazardous, as drivers overestimate the system's capabilities and are unable to react appropriately \cite{Nordhoff.2022}.

\subsection{Acceptance and Trust}
While acceptance according to \cite{Davis.1989} describes whether or why a technology is used or not, trust describes the attitude towards the technology without including its use \cite{Korber.2019}. 

Current studies on acceptance and trust in the context of automated driving focus in particular on which factors influence acceptance and trust of automated vehicles in society \cite{Cheng.2021,Haghzare.2021,Helgath.2018,Johansson.2021,Nastjuk.2020,Schmidt.2021,Weigl.2022,Wicki.2021,Winter.2022,Nordhoff.2020}.
The research examines how the actual experience of one \cite{Dillmann.2021, Haghzare.2021, Johansson.2021,Naujoks.2017,Nordhoff.2020,Schmidt.2021} or more trips \cite{Metz.2021,Zangi.2022} affects the development of acceptance and trust. 
The focus is often on takeover scenarios in SAE Level~3 automated vehicles, where responsibility for \gls{oedr} switches from the \gls{ads} to the human driver\cite{Dillmann.2021,Korber.2018,Gold.2015}.
Furthermore, the influence of acceptance and trust on driving behavior is currently under examination \cite{Korber.2018, Zangi.2022}.
Current experiments are mainly conducted in simulators \cite{Dillmann.2021,Haghzare.2021,Korber.2018,Metz.2021,Naujoks.2017,Schmidt.2021,Zangi.2022, Gold.2015}, or utilizing a \gls{woz} approach \cite{Nordhoff.2020,Johansson.2021} on a test track. 
Other studies are restricted to examining the society's acceptance and trust, both from a drivers and an outside perspective, through online surveys without considering actual experience \cite{Cheng.2021,Helgath.2018,Nastjuk.2020,Weigl.2022}.

Research findings reveal that experiencing automated driving firsthand through simulated rides increases both acceptance \cite{Metz.2021} and trust \cite{Gold.2015}, but that driving without experiencing the system's limitations quickly leads to an overestimation of the system's capabilities and thus to careless behavior \cite{Nordhoff.2022,Metz.2021}. 
Experiencing system limits, for example in the form of unexplained takeover requests, does however not lead to a decrease in acceptance \cite{Korber.2018}. 
Research indicates that individual factors, such as gender \cite{Weigl.2022}, age \cite{Haghzare.2021}, general attitudes towards technological innovations \cite{Nastjuk.2020}, as well as people's perceived usefulness and risk perceptions \cite{Nastjuk.2020}, influence acceptance and trust.
Another factor that affects acceptance and trust is the prevailing social norm. 
Inexperienced individuals may be particularly susceptible to the influence of the opinions of third parties in shaping acceptance \cite{Nastjuk.2020}. 
Media reporting on accidents involving automated vehicles also has influence on the acceptance and trust of individuals \cite{Wicki.2021}. 
Not least for that reason, according to the study by \cite{Cheng.2021}, one of the most common reasons for lack of acceptance is safety concerns.
Furthermore, the design of the \gls{ads} influences acceptance and trust. Additionally, the velocity and driving behavior of the \gls{ads} have an important influence on the acceptance of and trust in automated vehicles \cite{Haghzare.2021,Johansson.2021,Nordhoff.2020}.

\subsection{Aim of the Study}
In the German traffic environment, until the end of 2022, only series systems up to SAE Level 2 automation had road approval. 
However, the Mercedes-Benz Drive Pilot is the first SAE Level~3 \gls{ads} in series production to be approved for German road use according to UN-R157~\cite{UN-R157}.
Prior, only the Honda SENSING Elite has had a limited SAE Level~3 approval for Japan~\cite{honda_sensing}.
This has changed the traffic environment and highlighted a critical point of interest for the field of \gls{ads} research. 
Current methods, such as driving simulators or \gls{woz} tests, where a hidden human controls part of the \gls{ads} that appears to be automated, provide informative insights on the interaction between humans and \gls{ads}. 
However, they do not fully encapsulate the human interaction with live \gls{ads} in a real-world traffic environment. 
In particular, studies conducted in simulators or using \gls{woz} concepts show significant variation in the factors that influence acceptance and trust, such as risk perception \cite{Nastjuk.2020}, driving style, and velocity \cite{Haghzare.2021,Johansson.2021,Nordhoff.2020} compared to actual driving in real traffic. 
Therefore, it is imperative to validate the results described above through real traffic studies \cite{Frison.2020}.
The launch of the Drive Pilot enables researchers to conduct evaluations in the final \gls{odd} of the SAE Level~3 \gls{ads}, rather than in controlled or simulated environments. 
This provides a new opportunity to study human-automation interactions in the context of everyday use and real-world traffic conditions. \\
Therefore, this study will evaluate the development of acceptance and trust in the Level~3 Drive Pilot system, with a particular focus on drivers' first contact with Level~3 \gls{ads}. 
It will further address self-reported satisfaction and the emotional experience of using the system.
During the study, the following research questions will be answered by testing the EQS with participants who have no prior driving experience with automated driving: 

\begin{enumerate}
  \item How does the first contact with an SAE Level~3 \gls{ads} in the form of driving in real traffic affect trust in automated driving systems?
  \item How does the first contact with an SAE Level~3 \gls{ads} in the form of driving in real traffic affect the acceptance of automated driving systems?
  \item How is usability and user experience rated after the first contact with an SAE Level~3 \gls{ads}?
\end{enumerate}

\section{Methods}

The following delineates the method used to quantify the effect of first contact with \gls{ads} on the constructs of acceptance, trust, usability, and user experience.
For this, the dependent variables that will be quantified to describe the constructs are outlined. For better readability and to better distinguish the constructs and quantified variables from the abstract concepts they represent, the \quantb{constructs} and quantified \quant{dependent variables} are highlighted.
Further, the study environment, procedure, and participants are described to give insight into how the quantified variables were obtained.

\subsection{Dependent Variables}
\quantb{Acceptance} and \quantb{Trust} of the participants towards the \gls{ads} are evaluated as dependent variables. 
Additionally, \quant{Usability} and \quant{User Experience} are recorded.

To assess trust in automated systems, the German translation \cite{Pohler.2016} of the Scale of Trust \cite{Jian.2000} was used. 
This questionnaire represents the current standard for evaluating \quantb{Trust} in the context of interaction with automated vehicles \cite{Frison.2020} (Frison et al. 2020) and includes the subscales \quant{Trust} and \quant{Mistrust} as independent dimensions of trust. 
It consists of twelve items that participants rate on a 7-point Likert scale that ranges from 1 ("strongly disagree") to 7 ("strongly agree").

\quantb{Acceptance} is measured with the questionnaire according to \cite{Hewitt.2019}, which is an adaptation of the UTAUT \cite{Venkatesh.2003} and the CTAM \cite{Osswald.2012} for automated vehicles. 
The questionnaire includes the scales of the Autonomous Vehicle Acceptance Model: \quant{Performance Expectancy}, \quant{Effort Expectancy}, \quant{Attitude Toward Technology}, \quant{Social Influence}, \quant{Facilitating Conditions}, \quant{Self-Efficacy}, \quant{Anxiety}, \quant{Behavioral Intention}, and \quant{Perceived Safety}. 
The survey contains 26 items that are rated on a 7-point Likert scale from 1 ("strongly disagree") to 7 ("strongly agree"). 
Additionally, the importance of \quant{Hands}, \quant{Feet}, and \quant{Eyes} as method of control for the \gls{ads} is assessed on a similar 7-point scale.

To investigate self-reported satisfaction, the study utilizes the System Usability Scale (SUS) as described in \cite{Brooke.1996} \cite{Frison.2020}.
The SUS consists of 10 items, rated on a 5-point Likert scale ranging from 1 ("strongly disagree") to 5 ("strongly agree"). After analysis, the \quantb{System Usability} is displayed on a scale from 0 to 100. 
To also assess the emotional aspect of \quantb{User Experience} of system use, the \gls{ueq} \cite{Laugwitz.2008} is assessed.
This includes the six scales \quant{Attractiveness}, \quant{Perspicuity}, \quant{Efficiency}, \quant{Dependability}, \quant{Stimulation}, and \quant{Novelty} with a total of 26 opposing adjective pairs rated on a 7-point scale from -3 to 3.

\quant{Affinity for Technology} and \quant{General Trust} in technology are also recorded as confounding variables. 
\quant{Affinity for Technology} is measured using the German ATI scale according to \cite{Franke.2019}. 
It consists of 9 items, which the participants rate on a 6-point Likert scale from 1 ("completely disagree") to 6 ("completely agree"). 
The three items "One should be careful with unknown automated systems", "I trust a system rather than mistrust it" and "Automated systems generally work well" from the questionnaire of \cite{Korber.2019} were used to measure \quant{General Trust} in technology. 
These are rated on a 5-point Likert scale ranging from 1 ("strongly disagree") to 5 ("strongly agree").

\subsection{Study Environment}
\begin{figure}[h]
\centering
\includegraphics[width=1\linewidth]{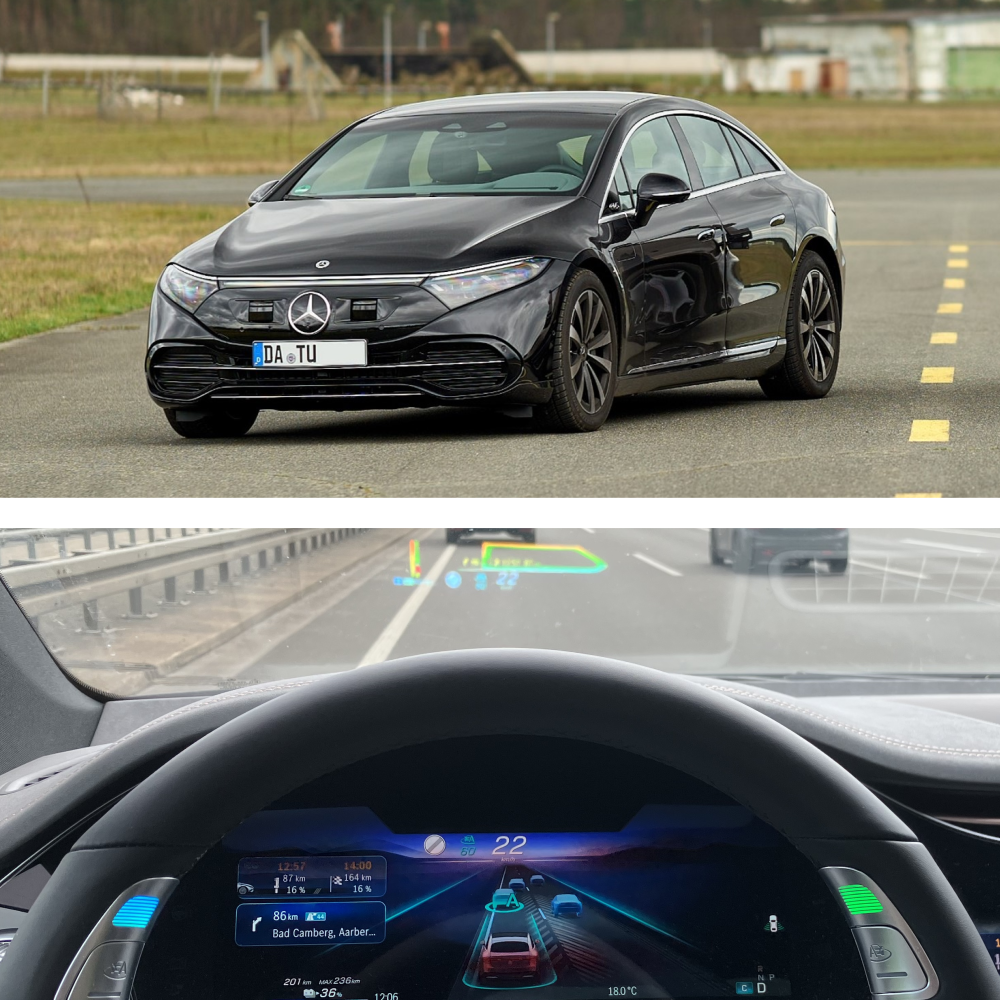} 
\caption{(Top) Used vehicle for study ( Mercedes-Benz
EQS (V297), (bottom) visual Human-Machine-Interface for the SAE Level 3 Drive Pilot while in operation.}
\label{fig:eqs}
\end{figure}

\begin{figure}[h]
\centering
\includegraphics[width=0.95\linewidth]{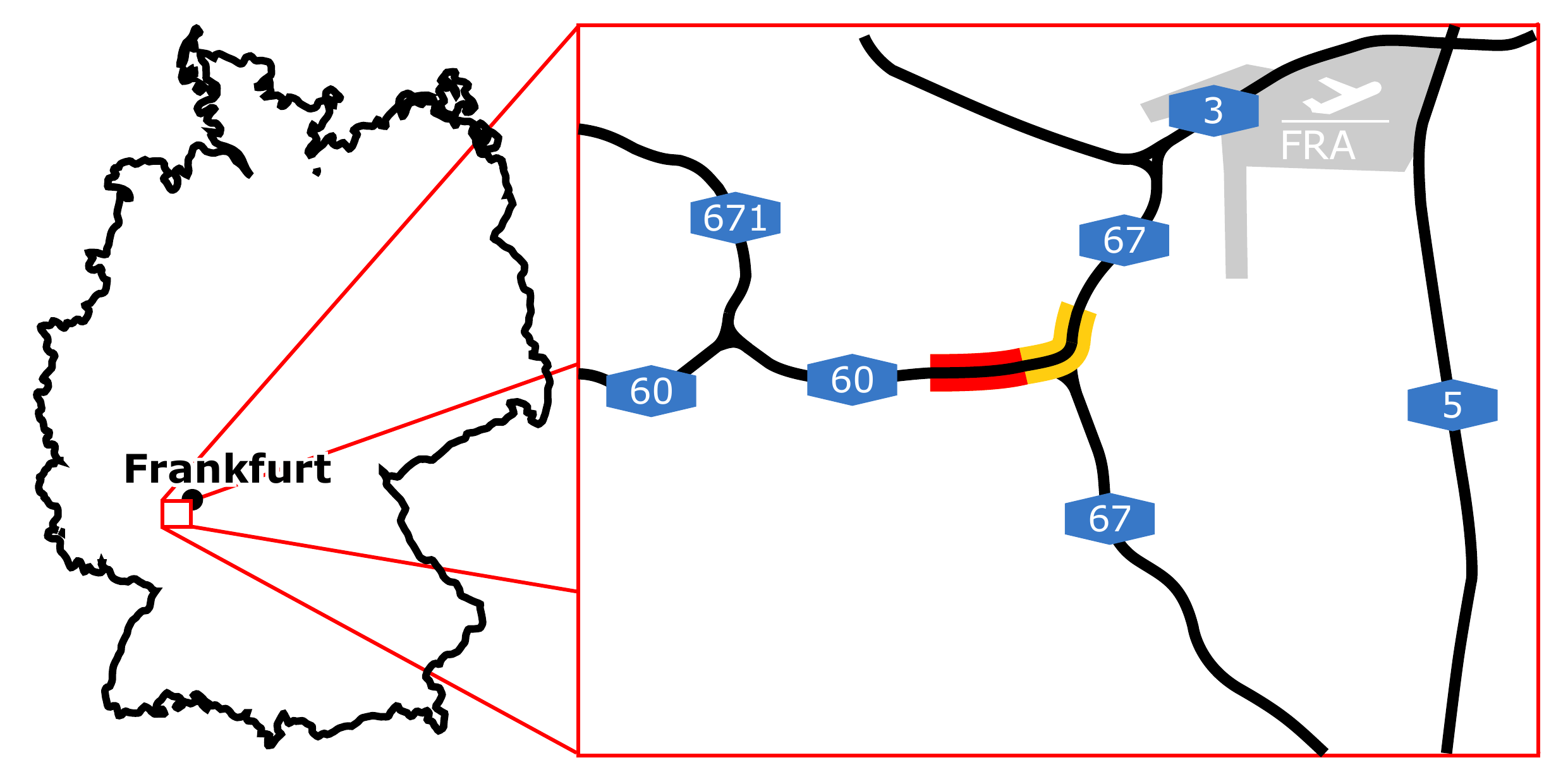} 
\caption{Map of Study Route. \textcolor{red}{Red} designates the used section of highway for \gls{ads} activation, while \textcolor{orange}{yellow} is the auxiliary section for creating a circular route.}
\label{fig:map}
\end{figure}

For the study, the subject vehicle was a Mercedes-Benz EQS (V297), and the subject \gls{ads} was the enabled SAE Level~3 Drive Pilot. 
The Drive Pilot enables automated driving on German highways up to a speed of 60 km/h and outside construction zones, while another vehicle is driving in front.
To enable activation, generally good environmental conditions have to be given.
For a more detailed description of the \gls{odd} see~\cite{mercedes_odd}.
Figure~\ref{fig:eqs} displays both the EQS and the HMI of the Drive Pilot. 
To maintain consistency throughout all trials, the driving within the study was carried out on a designated section of German highway during specific times of the day.
The data collection occurred on the final eastward section of A60, depicted in figure~\ref{fig:map}. 

Evaluations of the SAE Level~3 \gls{ads} by subjects were completed only on the 4 km long highway segment marked in red. 
Furthermore, a yellow-highlighted portion was utilized for making directional changes through the succeeding ramp, permitting smooth re-entry into traffic flow for subsequent rounds.
The full study round course is 17 km in length and enables two passes of the evaluation segment.
Each participant completed the round course at least once.
The 25 km long route up to the designated section was used to familiarize the participants with the vehicle. 
The study was conducted between May 8th and October 27th, during two intervals from 06:00 to 10:00 and from 15:00 to 19:00. 
The selected time frames matched commonly anticipated commuter traffic patterns, thus reflecting a realistic scenario in which end-users could engage an \gls{ads}. 
This scheduling permitted observing the performance of the \gls{ads} under varying levels of traffic density, which enhances the applicability of the study to everyday driving situations.
On average the \gls{ads} was activated a little more than five times per participant.
The average active duration of the \gls{ads} was 3.2 minutes with the maximum at 28 minutes and the minimum at around 30 seconds.

\subsection{Procedure}
\begin{figure}[h]
\centering
\includegraphics[width=1.0\linewidth]{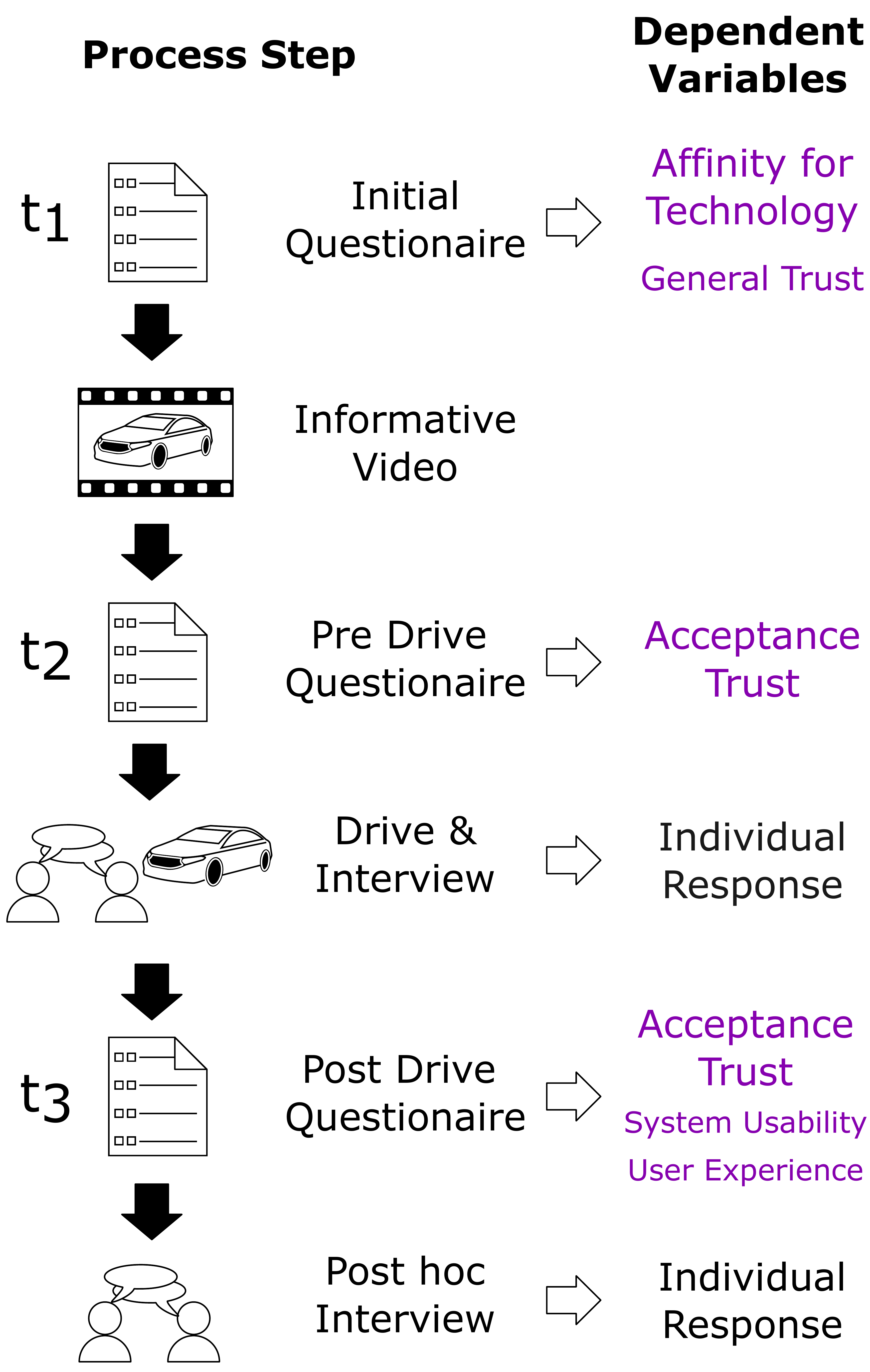} 
\caption{Flowchart for Study Procedure.}
\label{fig:process}
\end{figure}

The procedure of the study is shown in figure~\ref{fig:process}. 
Before driving, the participants' socio-demographic data, experience with SAE Level 1 and 2 \gls{ads}, affinity for technology, general trust in technology, and prior knowledge of automated vehicles are recorded at time $t_1$. 
This is followed by a 2-minute video demonstrating the functionality and handling of the SAE Level~3 \gls{ads} in the form of the Drive Pilot system of the Mercedes-Benz EQS, and answering questions about the system, as studies have shown that first encounters with automated driving can lead to inadequate driver performance due to lack of system knowledge \cite{Forster.2020}. 
To capture the impact of the first contact with the SAE Level~3 \gls{ads} on \quantb{Acceptance} and \quantb{Trust}, participants completed pre-journey ($t_2$) and post-journey ($t_3$) surveys regarding both dimensions. 
Meanwhile, the study supervisor both filmed and documented the subjects' reactions during the ride. 
After the ride, a post-hoc interview was conducted to collect further qualitative data on the influence of the first contact with the \gls{ads} on \quantb{Acceptance} and \quantb{Trust}. 
The participants were asked about their experience while driving, their ability to understand the behavior of the \gls{ads}, the usefulness of the feature, their comfort in distracting themselves from the driving task, and whether the vehicle would increase road safety. 
Since usability and user experience can both only be evaluated properly after the SAE Level~3 \gls{ads} has been experienced, the subjects answered both the SUS and the \gls{ueq} at time $t_3$ after driving.

\subsection{Participants}
\begin{figure*} [t]
\centering
\includegraphics[width=1\linewidth]{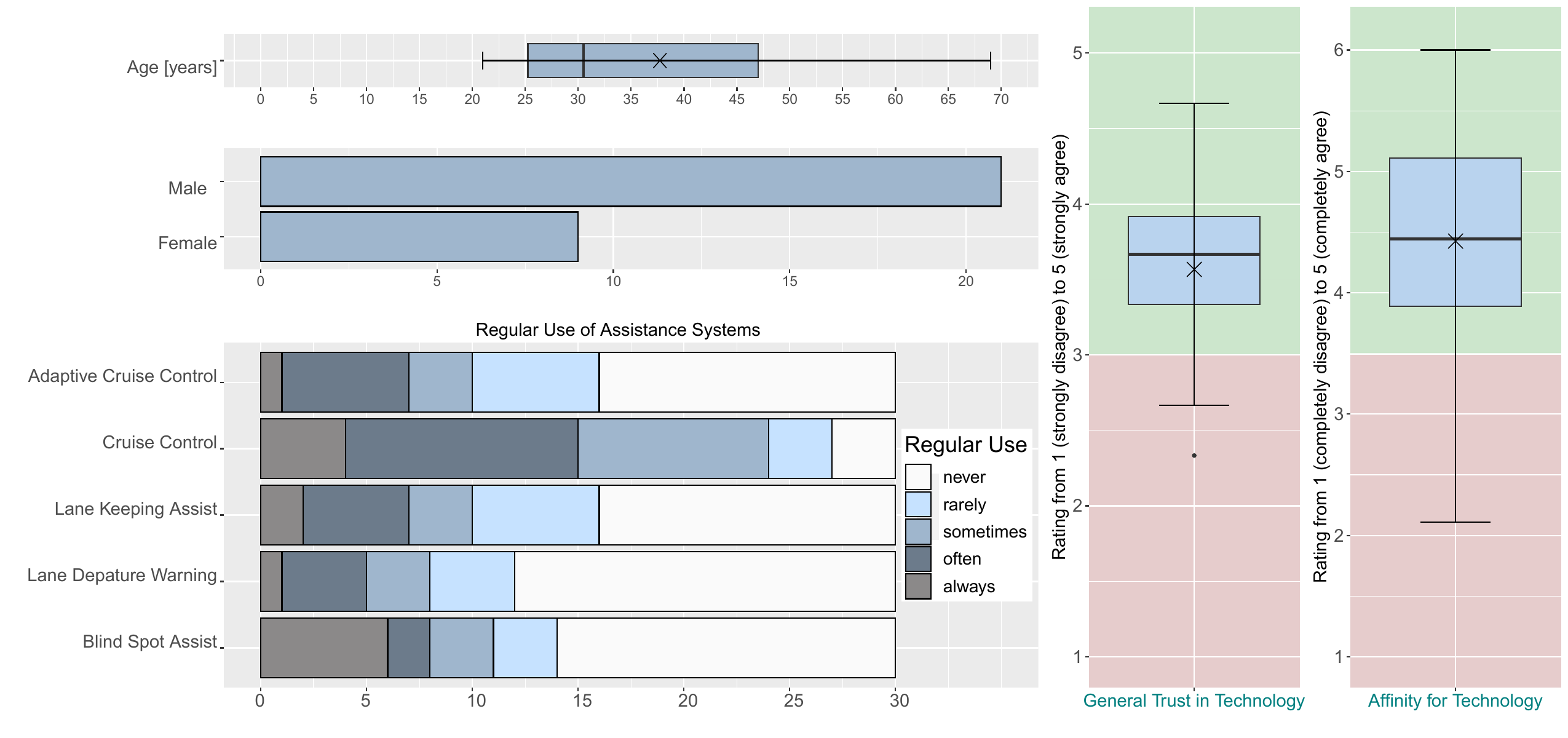} 
\caption{ Characteristics of the 30 participants: Age, Sex, Regular Use of Assistance Systems as well as General Trust in Technology and Affinity for Technology.}
\label{fig:Participants}
\end{figure*}
Participants were selected from an existing participant pool from institutes at TU Darmstadt, providing a foundational demographic. 
A key inclusion criterion for the study was participants' lack of professional experience with SAE Level~3 \gls{ads}. 
Experts with advanced knowledge and familiarity in this area were excluded, as their perspectives and behaviors may not accurately reflect those of the general public without experience in this field. 
In addition, the participants were required to have no prior SAE Level~3 \gls{ads} driving experience. 
Thirty participants were selected to take part in the study. 
The age of the 21 males and 9 females ranged from 21 to 69 years with a mean age of 37.7 years~(SD = 14.9). 
While most of them regularly use Cruise Control while driving, the Lane Keeping Assist, Lane Departure Warning, Adaptive Cruise Control and Blind Spot Assists are rarely used. 
Figure~\ref{fig:Participants} displays the characteristics of all 30 participants.
The axes for \quant{General Trust in Technology} and \quant{Affinity for Technology}, as well as all other applicable axes, have a colored background to signify positive and negative responses towards \gls{ads} or the Drive Pilot in particular.
As the figure further indicates, both the participants' \quant{General Trust in Technology} (M = 3.57, SD = 0.52) as well as their \quant{Affinity for Technology} (M = 4.43, SD = 0.94) were rather high.
Although the selection criteria were stringent, it is acknowledged that the participant pool remains subject to bias. 
A majority of the participants came from the Rhine-Main Area, with a notable concentration from Darmstadt. 
Darmstadt, also known as the "city of science," is characterized by a distinguished academic background and a robust high-tech industrial focus. 
This aspect inevitably shapes the demographics and cultural inclinations of its residents. Notably, the study's sample is dominated by younger participants due to the disproportionately young demographic of the city.

\section{Results}
To evaluate whether the first contact with SAE Level~3 \gls{ads} in the form of driving in real traffic caused a significant impact, the shift in mean for \quantb{Acceptance} and \quantb{Trust} is observed.
For this purpose, the data from questionnaires from $t_2$ to time point $t_3$ are used.
Using each corresponding pair of data samples points, represented by the mean $M$ and standard deviation $SD$, a mean shift with effect size $d=\frac{M_1-M_2}{SD_{\mathrm{pooled}}}$ is hypothesized.
For this work the pooled standard deviation $SD_{\mathrm{pooled}}$ is simplified to $\sqrt{\frac{1}{2}(SD_1^2+SD_2^2)}$.
To test this hypothesized effect, a t-test for paired samples is calculated.
For a test power of $1-\beta=0.95$ the resultant t-statistic $t$ and achieved significance level $p$ are presented.

\subsection{\quantb{Acceptance}}
\begin{figure*}[t]
\centering
\includegraphics[width=1\linewidth, clip]{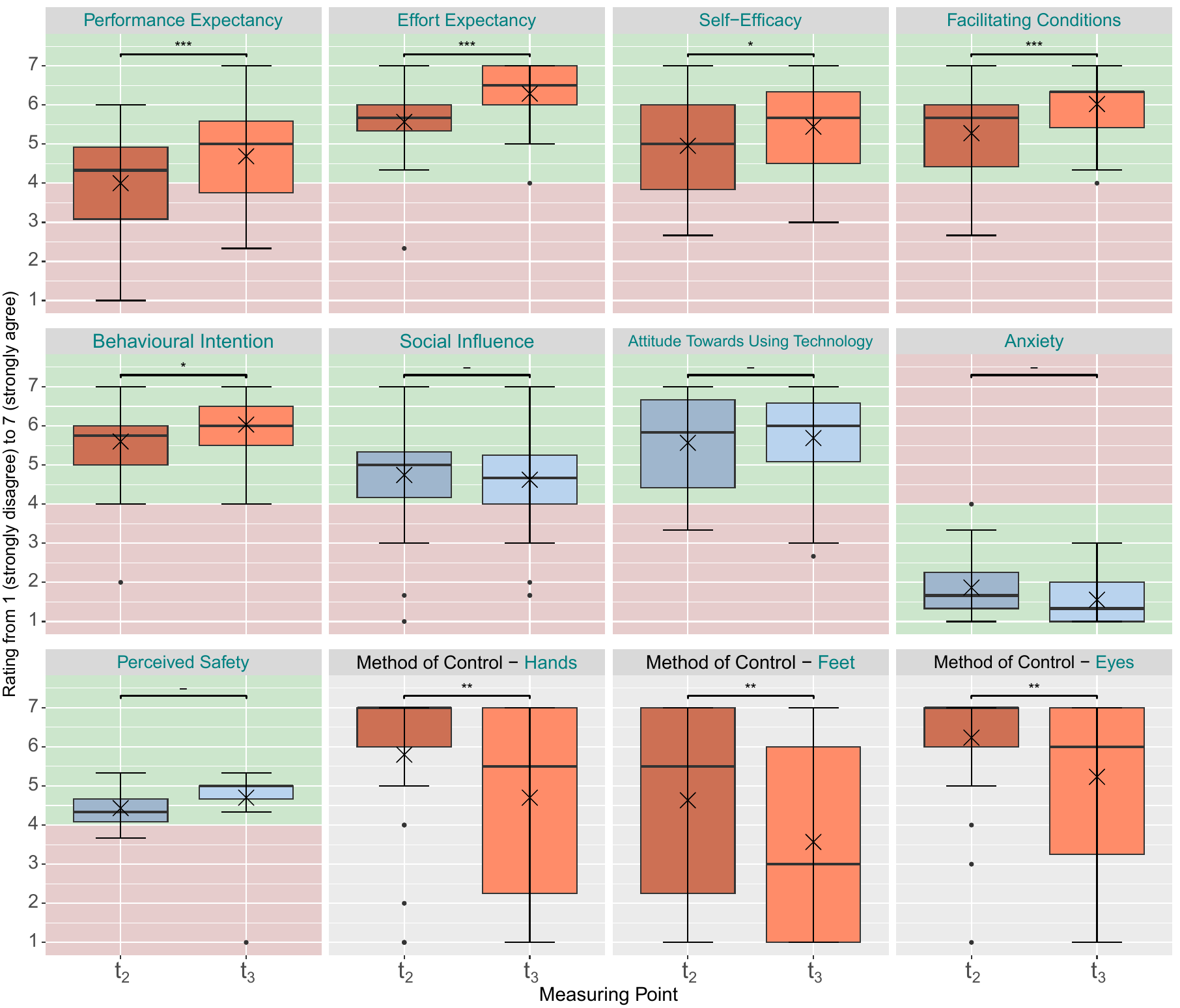}
\caption{Scales of \quantb{Acceptance} measured at $t_2$ and $t_3$ (*** significance level $<$ 0.001,  ** significance level $<$ 0.01, * significance level $<$ 0.05).}
\label{fig:acceptance}
\end{figure*}

Figure~\ref{fig:acceptance} displays the outcomes of the questionnaire regarding the factors of the Autonomous Vehicle Acceptance Model \cite{Hewitt.2019}. 
Blue box charts identify variables where no significant effect was observed, while red identifies those with a significant effect.

These results indicate that initial exposure to driving in real traffic significantly affects the \quant{Performance Expectancy} ($t(30) = -8.49, p < .001$), \quant{Effort Expectancy} ($t(30) = -3.98, p < .001$), \quant{Facilitating Condition} ($t(30) = -4.50, p < .001$) as well as \quant{Self-Efficacy} ($t(30) = -2.12, p = 0.043$) and \quant{Behavioral Intention} to use the vehicle ($t(30)=-2.06, p = 0.048$).
The mean value of \quant{Performance Expectancy} prior to first contact is $4.00~(SD=1.27)$, which increases to $6.29~(SD=0.76)$ afterwards. 
The mean \quant{Effort Expectancy} before first contact is $5.57~(SD=0.97)$, which also increases to $6.29~(SD=0.76)$ afterwards. 
Before first contact, the mean value for \quant{Facilitating Condition} is $5.28~(SD=1.08)$ and after contact, it is $6.02~(SD=0.78)$. 
Additionally, the mean value for \quant{Self-Efficacy} is $4.96~(SD=1.33)$ before the first contact and increases to $5.44~(SD=1.17)$ afterwards. 
The \quant{Behavioural Intention} to use the vehicle prior to first contact is an mean of $5.60~(SD = 1.05)$ and increases to $6.03~(SD = 0.72)$ thereafter. 
\quant{Performance Expectancy} has an effect size of $d = 1.477$, \quant{Effort Expectancy} of $d = .995$, \quant{Facilitating Condition}s of $d = .904$, \quant{Self-Efficacy} of $d = 1.264$, and \quant{Behavioral Intention} of $d = 1.150$, which corresponds to a strong effect in all cases, according to \cite{Cohen.1992}. 

Contrary, no significant affection was observed between the first contact with an SAE Level 3 \gls{ads} and \quant{Social Influence} ($t(30) = 0.65, p = 0.524$), \quant{Attitude Towards Using Technology} ($t(30) = -0.487, p = 0.630$), \quant{Anxiety} ($t(30) = 1.91, p=0.066$), and \quant{Perceived Safety} ($t(30) = -1.62, p =0.117$).
The mean value of \quant{Social Influence} prior to first contact is $4.74~(SD=1.37)$, which decreases to $4.62~(SD=1.33)$ afterwards.
For \quant{Attitude Towards Using Technology} the mean prior to first contact is $5.57~(SD=1.23)$, which afterwards increases to $5.69~(SD=1.21)$. 
Similarly, considering \quant{Anxiety} before the first contact is $1.87~(SD=0.71)$, and afterwards a decrease to $1.56~(SD=0.66)$ was observed. 
Lastly the mean value of \quant{Perceived Safety} prior to first contact is $4.43~(SD=0.41)$, which increases to $4.7~(SD=0.75)$ afterwards.

Additionally, the importance of \quant{Hands} ($t(30) = 3.66, p = .001$), Feet ($t(30) = 3.15, p = .001$), and \quant{Eyes} ($t(30) = 3.53, p = .001$) as methods of control for the \gls{ads} is rated significantly lower after the first contact with the SAE Level~3 \gls{ads}. 
The mean value for \quant{Hands} decreases from $5.80~(SD = 2.00)$ to $4.70~(SD = 2.44)$; for \quant{Feet}, it falls from $4.63~(SD = 2.41)$ to $3.57~(SD = 2.39)$, and for \quant{Eyes}, it decreases from $6.23~(SD = 1.43)$ to $5.23~(SD = 2.03)$. 
The effect size is $d = 1.647$ for \quant{Hands}, $d = 1.856$ for \quant{Feet}, and $d = 1.554$ for \quant{Eyes}, which corresponds to a strong effect in all cases, according to \cite{Cohen.1992}. 


\subsection{\quantb{Trust}}
\begin{figure}[h]
\includegraphics[width=0.9\linewidth]{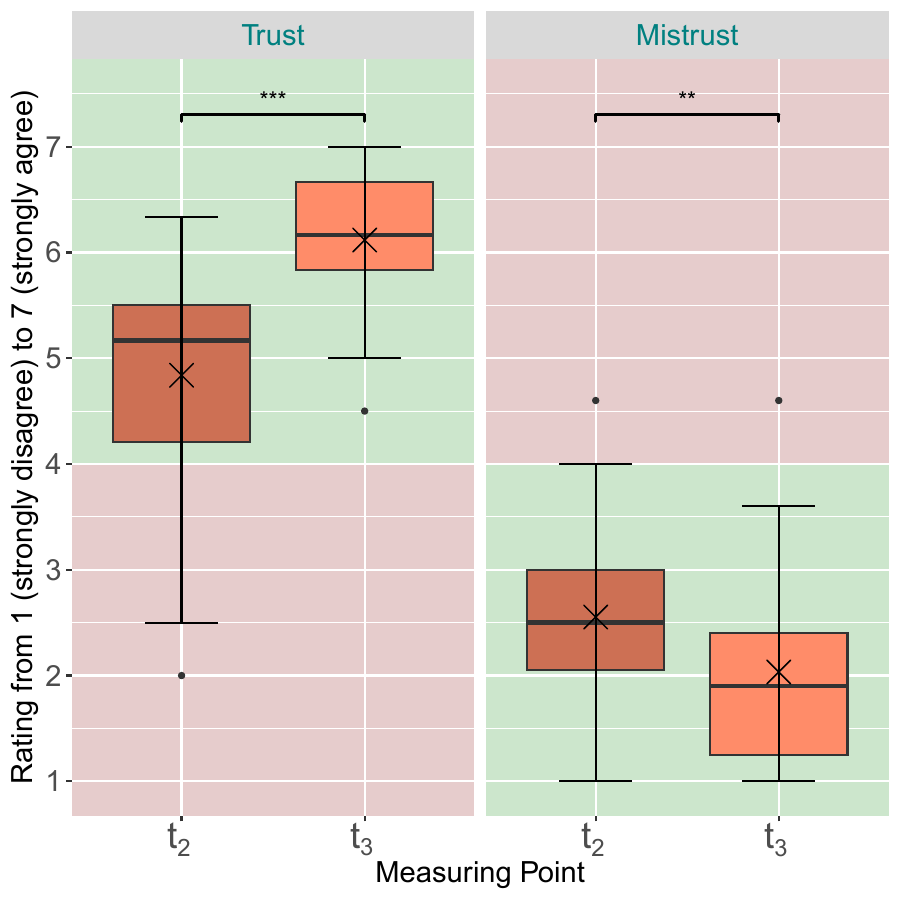}
\centering
\caption{Scales of \quantb{Trust} measured at $t_2$ and $t_3$ (*** significance level $< 0.001$,  ** significance level $< 0.01$, * significance level $< 0.05$).}
\label{fig:trust}
\end{figure}

As figure~\ref{fig:trust} shows, the mean level of \quant{Trust} before initial contact ($t_2$) was $4.84~(SD = 1.08)$ and increased to $6.12~(SD = 0.6)$ after the exposure ($t_3$). 
Meanwhile, the mean level of \quant{Mistrust} before initial contact was $2.55~(SD = 0.79)$ and decreased to $2.03~(SD = 0.93)$ after the exposure. 
The results demonstrate that the first contact, in the form of driving in real traffic, has a significant effect on both \quant{Trust} ($t(30) = -6.2, p < .001$) and \quant{Mistrust} ($t(30) = -3.209, p = .01$). 
The effect size for \quant{Trust} and \quant{Mistrust} is $d = 1.13$ and $d = .887$, which corresponds to a strong effect in both cases, according to \cite{Cohen.1992}.

\subsection{\quantb{Usability} and \quantb{User Experience} }
\begin{figure*}[t]
\includegraphics[width=0.9\linewidth]{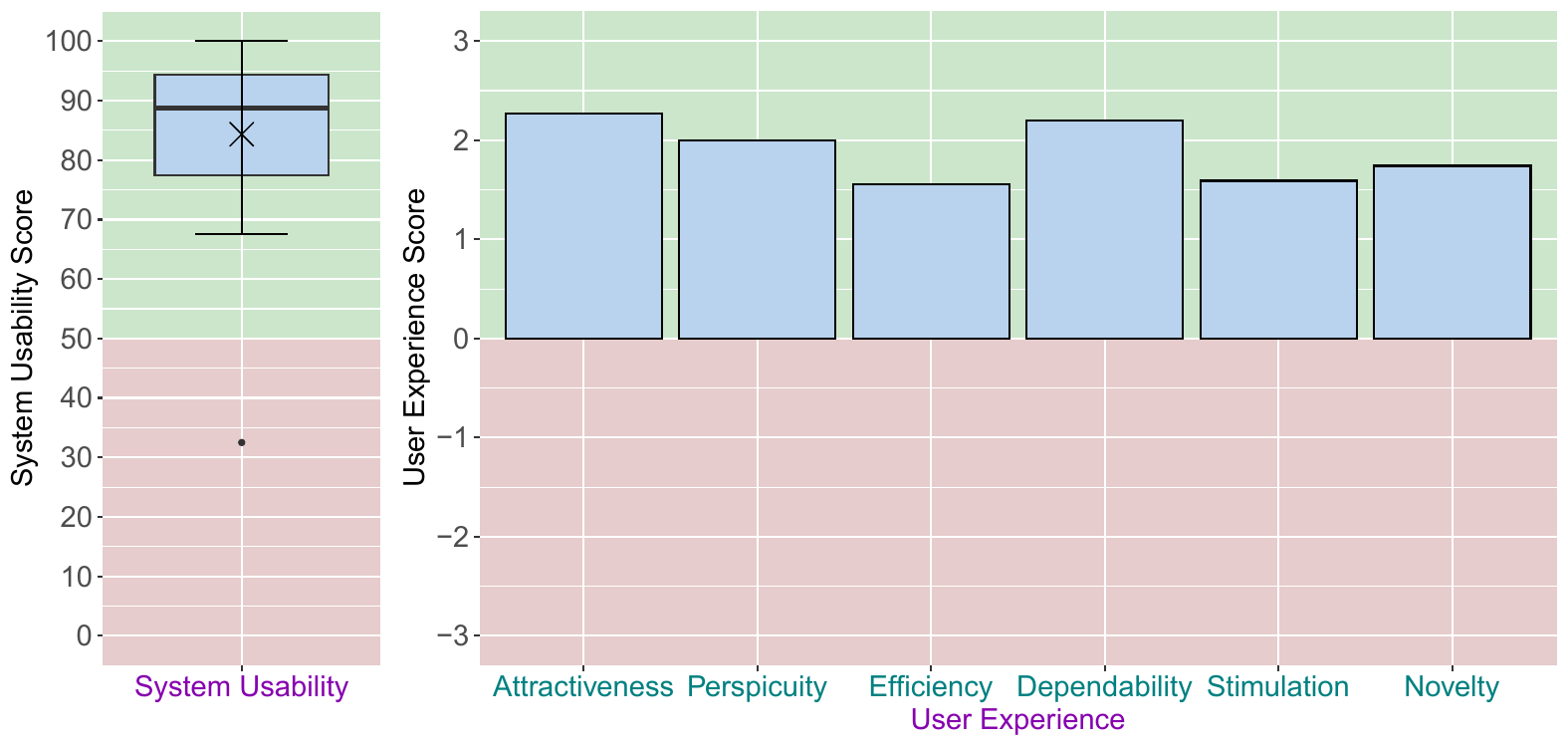}
\centering
\caption{\quantb{System Usability} and \quantb{User Experience} measured at $t_3$.}
\label{fig:UX}
\end{figure*}

The results of the System Usability Scale \cite{Brooke.1996} presented in figure~\ref{fig:UX} show that \quantb{System Usability} is rated as good to excellent $(M = 84.33, SD = 13.52)$ \cite{Bangor.2008}. 
Figure~\ref{fig:UX} also presents the \quantb{User Experience} results of the \gls{ueq} \cite{Laugwitz.2008}, which exhibit good to very good results on all scales, particularly positive ratings for \quant{Attractiveness} $(M = 2.27, SD = 0.56)$ and \quant{Dependability} $(M = 2.20, SD = 0.63)$.

\subsection{Posthoc  Interviews}
The post-hoc interview was analyzed according to the procedure for qualitative content analysis of \cite{Mayring.2022}. 
In sum, all participants felt safe during the automated journey, although ten of them expressed feeling safe during the automated journey only after a short familiarization period. 
The majority of participants understood the \gls{ads} behavior and the driving style was deemed highly anticipatory. 
However, the reasons why the \gls{ads} deactivates itself or refused to activate again could not always be understood. 
For instance, participants who received the message "Road section unsuitable" did not understand why it was unsuitable. 
Nine participants felt unable to disengage from the driving task because they believed that they needed to monitor the \gls{ads}, but assumed that it would be possible to disengage after a longer period of familiarization. 
The dominant opinion of the participants was that regular use of the \gls{ads} would increase trust and lead to the ability to turn away from the driving task more often. 
As a result, the entire driving experience would be more relaxed and restful. Overall, study subjects believed that \gls{ads} would enhance road safety due to their anticipatory driving style and ability to provide a better overview of traffic conditions compared to human drivers. 
However, five individuals who participated in the study expressed concern about the confusion between the SAE Level~2 and the SAE Level~3 \gls{ads} and the hazards that could be associated with such mode confusion when the SAE Level~3 ADS is deactivated. 
The ADS feature was deemed practical or even comfortable by 28 participants, although limitations such as low speed and early deactivation before road works were considered to be too restricting for regular use. 
On the other hand, 4 participants reported that using the \gls{ads} was just as strenuous as driving, as they had to keep an eye on the \gls{ads} and their surroundings.

\section{Discussion}
In this research, the influence of the first contact with the approved SAE Level~3 ADS in the form of the Drive Pilot in the Mercedes-Benz EQS in real traffic situations on \quantb{Trust}, \quantb{Acceptance}, \quantb{Usability}, and \quantb{User Experience} was investigated.
The study found a substantial increase in both \quantb{Acceptance} and \quantb{Trust} after first contact with the SAE Level~3 ADS. 
These outcomes validate prior simulator study results \cite{Metz.2021, Gold.2015}. 
Overall all observed quantities for \quantb{Acceptance}, \quantb{Trust}, \quantb{System Usability}, and \quantb{User Experience} exhibited a positive initial situation towards the \gls{ads}.
In particular, \quantb{Acceptance} of the \gls{ads} in terms of fast, cost-effective, and safe destination reach, as well as simplicity and ease of use, was rated significantly higher after the driving experience than expected before the experience. 
The study subjects rated their control of the vehicle, their knowledge of how to use the vehicle, and the feasibility of implementing the necessary technical and infrastructural requirements significantly higher after the driving experience than before. 
They also gave significantly higher ratings to their \quant{Self-Efficacy} in handling the vehicle and to their \quant{Intention} to use the vehicle after the trip.

Additionally, the findings of the simulator studies confirm that unexplained takeover requests or unclear reasoning for refusing to activate the ADS did not have a negative impact on \quantb{Trust} or \quantb{Acceptance} \cite{Korber.2018}. 
Furthermore, similar to previous studies \cite{Nordhoff.2022,Metz.2021}, tendencies for careless misuse was observed during SAE Level~3 \gls{ads} driving.  
For example, two participants attempted to change their seating position during automated driving, which may have inhibited or prevented them from quickly and adequately retaking the driving task.
It was also observed that the study subjects neglected the driving task after using the SAE Level~3 ADS, even when using the SAE Level~2 \gls{ads}, and accordingly let go of the steering wheel, causing the SAE Level~2 \gls{ads} to issue a warning to retake the steering wheel.
Whether this was due to overestimation \cite{Metz.2021,Nordhoff.2022} or rather due to mode confusion could not be conclusively determined.

Contrary to prior findings \cite{Nastjuk.2020}, this study could not establish a significant effect of general attitudes toward technological innovation on \quantb{Acceptance} and \quantb{Trust}. 
Nevertheless, it is worth mentioning that most participants tended to have a high \quant{Affinity for Technology} or a \quant{General Trust in Technology}. 
It is hypothesized that individuals who have an aversion to technological innovation might be less inclined to take part in studies on \gls{ads}. 
Thus, it is necessary to consider whether a refusal to participate is due to a general mistrust of technological innovation. 
While strict selection criteria were applied to select participants, it is recognized that the participant pool may not be free of bias. 
It should be noted that the study sample consisted mostly of younger participants. 
Although this provides important insights into the views of young populations on \gls{ads}, it may not accurately represent older populations who may interact differently with such systems, due to an under representation in the study participants.
For future studies, it is recommended to diversify the sample of participants and include people with a lower affinity for or trust in technology.
Research of interest could also include observing longer or multiple trips in real traffic to understand how driving behavior adapts, since participants mentioned that they would adapt their behavior after a longer period of familiarization. 
In this context, the use of methods to determine driver readiness would also be interesting. 
In addition, to avoid mode confusion, driver-vehicle communication needs to be improved.
A possible approach for this is taking into account additional driver information, like analyzing eye movements while driving.

\section{Conclusion}
In summary, this study showed that the participants' first experience as a driver with an SAE Level~3 \gls{ads} was generally positive, increasing their acceptance and trust of the \gls{ads}. 
While participants acknowledge that the realization of automated functions represents a major achievement and has great potential, they also point out the need to further expand the limitations of the \gls{ads} to maximize the benefits for the driver, as well as to improve communication between the vehicle and the driver to avoid mode confusion and misuse.

In terms of introduction strategies for \gls{ads}, the results indicate, that an iterative approach, where the \gls{odd} of the \gls{ads} is being expanded has benefits over approaches, where the goal is to initially release the \gls{ads} with a large \gls{odd}.
Incremental releases enable early first contacts, which as this study shows, have a positive effect on acceptance and trust.

\section*{Acknowledgments}
The authors thank all participants, and especially Yannick Schwarz, Jonas Lehmann and Meho Mahalbasic for their support in executing this study.

\section*{Declaration of Conflict of Interest}
The author Prof. Peters was not involved in the development of the Drive Pilot during his prior employment with Mercedes-Benz Group AG. 
The EQS was ordinarily purchased by the Technical University of Darmstadt.
Therefore the Authors declare that there is no conflict of interest.


\bibliographystyle{IEEEtran}
\bibliography{literature}



\section{Biography Section}

\vspace{-33pt}
\begin{IEEEbiography}
[{\includegraphics[width=1in,height=1.25in,clip,keepaspectratio]{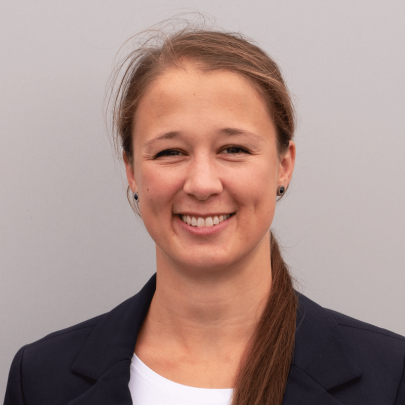}}]{Sarah Schwindt-Drews} Sarah Schwindt-Drews studied mechanical engineering at the Technical University of Darmstadt, where she received the M.Sc. in 2020. 
Since 2020 she is a research associate at the Institute for Ergonomics and Human Factors at Technical University of Darmstadt and working towards the Ph.D. degree. Her research interests center around the mental models of drivers in automated vehicles.
\end{IEEEbiography}

\vspace{-33pt}
\begin{IEEEbiography}[{\includegraphics[width=1in,height=1.25in,clip,keepaspectratio]{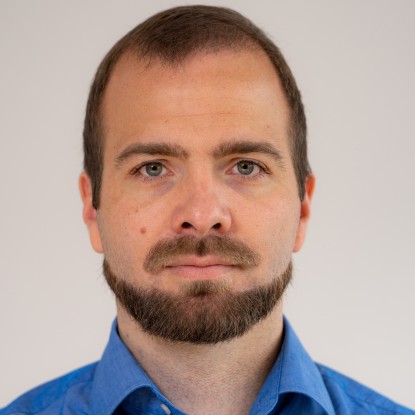}}]{Kai Storms}
studied Mechanical and Process engineering at Technical University of Darmstadt, where he received the M.Sc. in 2020. He is currently a research associate at Technical University of Darmstadt, Germany and working towards the Ph.D. degree. Since 2020, he has been conducting research within highly automated driving. His research interest include verification and validation of automated vehicles, with a focus on data reduction.
\end{IEEEbiography}

\vspace{-33pt}
\begin{IEEEbiography}
[{\includegraphics[width=1in,height=1.25in,clip,keepaspectratio]{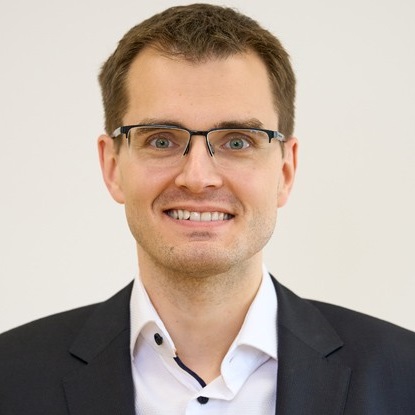}}]{Steven Peters} was born in 1987, received his PhD (Dr.-Ing.) in 2013, at Karlsruhe Institute of Technology, Karlsruhe, Baden-Württemberg, Germany.
From 2016 to 2022 he worked as Manager of Artificial Intelligence Research at Mercedes-Benz AG in Germany. He is a Full Professor at the Technical University of Darmstadt, Darmstadt, Germany and heads the Institute of Automotive Engineering in the Department of Mechanical Engineering since 2022.
\end{IEEEbiography}

\vspace{-33pt}
\begin{IEEEbiography}
[{\includegraphics[width=1in,height=1.25in,clip,keepaspectratio]{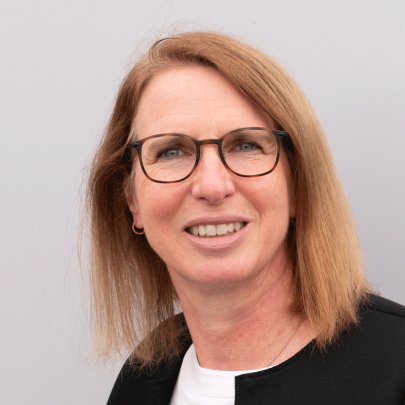}}]{Bettina Abendroth} studied industrial engineering, majoring in mechanical engineering at the TU Darmstadt University of Technology. She completed her Ph.D. in July 2001. Her topic was driver types of vehicles and driving support at longitudinal guiding. She is the deputy director of the Institute of Ergonomics and Human Factors and heads the Human-Machine Interaction \& Mobility research group.
\end{IEEEbiography}

\vfill

\end{document}